\documentclass[letter,twocolumn]{jpsj3}
\usepackage{txfonts, color}

\title{Zero-Field Miniature Skyrmion Crystal and Chiral Domain State in Breathing-Kagome Antiferromagnets}

\author{Kazushi Aoyama$^1$\thanks{jpsj{\_}aoyama@ess.sci.osaka-u.ac.jp} and Hikaru Kawamura$^2$}
\inst{$^1$Department of Earth and Space Science, Graduate School of Science, Osaka University, Osaka 560-0043, Japan \\
$^2$Molecular Photoscience Research Center, Kobe University, Kobe 657-8501, Japan} %\\

\abst{The stability of a miniature skyrmion crystal (SkX) with only a small number of spins in the magnetic unit cell has been theoretically investigated in $J_1$-$J_3$ antiferromagnets on the breathing kagome lattice with a single-ion anisotropy $D$ at zero field. It is found by means of Monte Carlo simulations that due to the breathing bond-alternation, a zero-field triple-${\bf Q}$ miniature SkX can be stabilized not only in the specific case of $D=0$ [K. Aoyama and H. Kawamura, Phys. Rev. B {\bf 105}, L100407 (2022)] but also in more general situations with easy-axis ($D<0$) and easy-plane ($D>0$) anisotropies which favor triple-${\bf Q}$ collinear and noncoplanar states, respectively. Since the SkX and anti-SkX each having positive or negative chirality are energetically degenerate, the topological Hall effect of alternative sign is possible at zero field. It is also found that reflecting the chiral degeneracy, the collinear and coplanar phases preempting the SkX phase possess random domain structures consisting of positive- and negative-chirality clusters.}

\begin{document}
\maketitle

%skyrmion --> chiral object --> rich variety of zero-field chiral orders 
A magnetic skyrmion is a noncoplanar swirling texture composed of many spins whose total solid angle is quantized in units of $4\pi$ to an integer value $n_{\rm sk}$ \cite{SkX_review_Nagaosa-Tokura_13}. Noting that a solid angle $\Omega_{ijk}$ subtended by three spins ${\bf S}_i$, ${\bf S}_j$, and ${\bf S}_k$ is calculated with the use of the scalar spin chirality $\chi_{ijk}={\bf S}_i\cdot({\bf S}_j\times{\bf S}_k)$ \cite{SolidAngle_OOsterom_83}, the skyrmion can be regarded as a chirality-related topological object. This inversely suggests that if a noncoplanar spin structure with $\chi_{ijk}\neq 0$ possesses a nonzero integer $n_{\rm sk}$, it might be understood as a kind of the skyrmion. Previously, we showed that $J_1$-$J_3$ Heisenberg antiferromagnets on the breathing kagome lattice host a zero-field chiral order whose spin structure can be viewed as a periodic array of small-size skyrmions \cite{KagomeSkX_AK_22}. In this work, we investigate the stability of this zero-field miniature skyrmion crystal against magnetic anisotropies which more or less exist in real materials.

%skyrmion and Z phase in an applied field
As is well known, in bulk magnets, the skyrmions are often stabilized in the form of their two-dimensional crystalline order, the so-called skyrmion crystal (SkX), where in most cases, the Dzaloshinskii-Moriya (DM) interaction and an applied magnetic field play essential roles for the SkX formation \cite{SkX_Bogdanov_89, SkX_Yi_09, SkX_Buhrandt_13, MnSi_Muhlbauer_09, MnSi_Neubauer_09, FeCoSi_Yu_10, FeGe_Yu_11, Fefilm_Heinze_11, Cu2OSeO3_Seki_12, CoZnMn_Tokunaga_15, GaV4S8_Kezsmarki_15, GaV4Se8_Fujima_17, GaV4Se8_Bordacs_17, VOSe2O5_Kurumaji_17, AntiSkX_Nayak_17, EuPtSi_Kakihana_18, EuPtSi_Kaneko_19}. Of recent particular interest is the SkX in the absence of the DM interaction \cite{SkX_Okubo_12, SkX_Leonov_15, SkX_Lin_16, SkX_top2_Ozawa_prl_17, SkX-RKKY_Hayami_17, SkX_Lin_18, SkX-RKKY_Hayami_19, SkX-RKKY_Wang_20, SkX-bondaniso_Hayami_21, SkX-bondaniso_Batista_21, SkX-RKKY_Mitsumoto_21, SkX-RKKY_Mitsumoto_22, Gd2PdSi3_Kurumaji_19, Gd2PdSi3_Paddison_prl_22, GdRuAl_Hirschberger_natcom_19, GdRu2Si2_Khanh_20, EuAl4_Shang_prb_21, EuAl4_Takagi_natcom_22}. Several mechanisms other than the DM interaction, e.g., competitions between exchange interactions (magnetic frustration) exemplified by the Ruderman-Kittel-Kasuya-Yosida interaction \cite{SkX_Okubo_12, SkX-RKKY_Mitsumoto_21, SkX-RKKY_Mitsumoto_22}, multi-spin interactions originating from the Fermi surface nesting in the conduction band \cite{SkX_top2_Ozawa_prl_17, SkX-RKKY_Hayami_17}, and anisotropic exchange interactions \cite{SkX-bondaniso_Hayami_21, SkX-bondaniso_Batista_21}, have been reported so far. In the DM-free systems, positive and negative chiralities are energetically degenerate, so that the SkX and anti-SkX each having positive/negative net chirality are equally possible. Reflecting the chiral degeneracy, a chiral domain state consisting of randomly-distributed SkX and anti-SkX domains, the so-called $Z$ phase \cite{SkX_Okubo_12, SkX-RKKY_Mitsumoto_21, SkX-RKKY_Mitsumoto_22}, can be realized as an equilibrium state distinct from the SkX and paramagnetic phases. Apart from these DM and non-DM mechanisms, the SkX's commonly appear in the applied magnetic field. Then, the question is whether or not the SkX physics can appear even in the absence of the applied field and a spontaneous uniform magnetization.

\begin{figure}
\begin{center}
\includegraphics[scale=0.69]{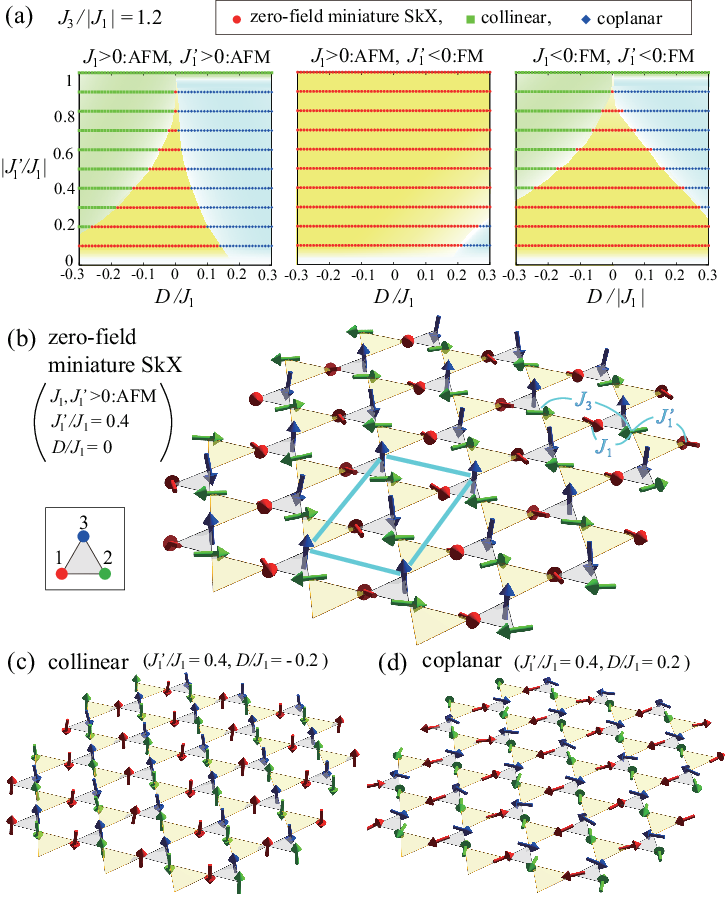}
\caption{(Color online) MC results obtained at $T/|J_1|=0.01$ for $J_3/|J_1|=1.2$. (a) The parameter dependence of the low-temperature spin structure, where the stability regions are determined mainly from the $L=18$ data. (b)-(d) MC snapshots of the miniature SkX, collinear, and coplanar states, which are, respectively, taken at $D/J_1=0$, $-0.2$, and $0.2$ for $J_1'/J_1=0.4$ with AFM $J_1>0$ and $J_1'>0$, where red, blue, and green arrows, respectively, represent spins on the corners of 1, 2, and 3 of the small triangle shown in the inset. In (b), a cyan rectangle indicates a single miniature skyrmion, and the definitions of $J_1$, $J_1'$, and $J_3$ are also shown.  
 \label{fig:Paradep_structure} }
\end{center}
\end{figure}

% brief summary of zero field chiral order
One reported example of such a zero-field SkX is a commensurate SkX on the triangular lattice induced by the Kondo coupling to conduction electrons \cite{SkX_top2_Ozawa_prl_17}, and spin-orbit-coupling mechanisms are recently reported as well \cite{SkX_Amoroso_natcom_20, SkX_Yambe_scirep_21}. Focusing on the aspect that the SkX is a uniform chiral order with nonzero integer $n_{\rm sk}$, however, we notice that several zero-field chiral orders with $n_{\rm sk} \neq 0$ have been theoretically studied: 4-sublattice \cite{4sub_Momoi_prl_97,4sub_Martin_prl_08,4sub_Akagi_jpsj_10,4sub_Akagi_prl_12} and 12-sublattice \cite{12subTopo_Barros_prb_14} orders on the triangular and kagome lattices, respectively, and a meron crystal on the square lattice \cite{SkX-bondaniso_Batista_21}. The origins of these orders with the topological spin structures are essentially the same as those listed above for the DM-free in-field SkX, i.e., the Fermi surface nesting, the multi-spin interaction, or the anisotropies in the exchange interactions. Recently, we demonstrated that the breathing bond-alternation of the lattice serves as another mechanism and that a zero-field chiral order emerging on the breathing kagome lattice possesses a SkX structure with 12 spins in its magnetic unit cell [see Ref. \cite{KagomeSkX_AK_22} and Fig. \ref{fig:Paradep_structure} (b)]. In this case, a single skyrmion involves only a fixed small number of spins on the discrete lattice sites [see the cyan rectangle in Fig. \ref{fig:Paradep_structure} (b)], so that we call it a miniature skyrmion to distinguish it from the conventional skyrmion involving many spins. In the miniature case, although the topological stability of the conventional skyrmion would be lost in the {\it single-excitation dynamics}, it remains present in the {\it crystal state} of our interest.

%explanation of the miniature SkX and analogy to hedgehog
The breathing kagome lattice consists of corner-sharing small and large triangles. The characteristic feature of this breathing lattice is that the nearest neighbor (NN) exchange interactions on small and large triangles $J_1$ and $J_1'$ take different values due to the bond-length difference. Previously, we showed that in $J_3$-dominant $J_1$-$J_3$ Heisenberg antiferromagnets, a zero-field chiral order having the miniature SkX/anti-SkX structure with $n_{\rm sk}=\pm 2$ becomes stable for $J_1 \neq J_1'$.
As the formation of the conventional in-field SkX is assisted by an additional easy-axis anisotropy \cite{SkX_Leonov_15, SkX_Lin_16}, the stability of the zero-field SkX should also be affected by the magnetic anisotropy. In this work, we examine effects of single-ion magnetic anisotropies on the zero-field miniature SkX. We will show by means of Monte Carlo (MC) simulations that the miniature SkX can be realized not only in the ideally isotropic case but also in more realistic cases with the magnetic anisotropies and that chiral domain states analogous to the $Z$ phase appear.

%model
The spin Hamiltonian we consider is given by
\begin{eqnarray}\label{eq:Hamiltonian}
{\cal H} &=& J_1 \sum_{\langle i,j \rangle_S} {\bf S}_i\cdot{\bf S}_j + J_1' \sum_{\langle i,j \rangle_L} {\bf S}_i\cdot{\bf S}_j + J_3\sum_{\langle \langle  i,j \rangle \rangle } {\bf S}_i\cdot{\bf S}_j \nonumber\\
&+& D\sum_{i}(S_i^z)^2,
\end{eqnarray}
where ${\bf S}_i$ is a classical spin, the summation $\sum_{\langle \rangle_{S(L)}}$ runs over site pairs on small (large) triangles having the NN interaction $J_1$ ($J_1'$), $J_3>0$ is the third NN antiferromagnetic interaction along the bond direction [see Fig. \ref{fig:Paradep_structure} (b)], and $D<0$ ($D>0$) represents an easy-axis (easy-plane) anisotropy. The DM interaction inherent to the kagome lattice is assumed to be negligibly small.
The ratio $J_1'/J_1$ characterizes the breathing lattice structure; $J_1'/J_1=1$ and $J_1'/J_1 \neq 1$ correspond to the uniform and breathing kagome lattices, respectively. For relatively strong $J_3$ at $D=0$, ordering vectors turn out to be ${\bf Q}_1=\frac{\pi}{2a}(-1,-\frac{1}{\sqrt{3}})$, ${\bf Q}_2=\frac{\pi}{2a}(1,-\frac{1}{\sqrt{3}})$, and ${\bf Q}_3=\frac{\pi}{2a}(0,\frac{2}{\sqrt{3}})$ with side length of each triangle $a=1$ (for simplicity, the bond-length difference is incorporated only in the nonequivalent $J_1$ and $J_1'$ and not in the real-space length scale). Such a $J_3$-dominant $J_1$-$J_3$ model is proposed for the uniform kagome antiferromagnet BaCu$_3$V$_2$O$_8$(OD)$_2$ \cite{CoplanarOct_Boldrin_prl_18}.
In the isotropic case of $D=0$, a 12-sublattice triple-${\bf Q}$ state characterized by ${\bf Q}_1$, ${\bf Q}_2$, and ${\bf Q}_3$ takes a noncoplanar or coplanar spin configuration for $J_1'/J_1 \neq 1$ \cite{KagomeSkX_AK_22}, whereas for $J_1'/J_1 = 1$, it takes a collinear configuration favored by thermal fluctuations \cite{RMO-collinear_Grison_prb_20}. The noncoplanar state corresponds to the zero-field miniature SkX. We note that single- and double-${\bf Q}$ states do not appear for the specific ordering vectors of ${\bf Q}_1$, ${\bf Q}_2$, and ${\bf Q}_3$ which are induced by large $J_3$ and are relatively robust against an additional $J_2$ \cite{KagomeSkX_AK_22}. In this work, we pick up $J_3/|J_1|=1.2$ in a large parameter space available for the miniature SkX (see Fig. 4 in Ref. \cite{KagomeSkX_AK_22}), and examine the $D$ dependence of the ordered state.

%MC simulation
To investigate ordering properties, we perform Monte Carlo simulations.  
In the MC simulations, $1\times 10^6$ - $2\times 10^6$ MC sweeps are carried out under the periodic boundary condition, and the first half is discarded for thermalization, where a single spin flip at each site consists of the conventional Metropolis update and a successive over-relaxation-like process in which we try to rotate a spin by the angle $\pi$ around the local mean field \cite{Loop_Shinaoka_14, Site_AK_19}. Observations are done at every MC sweep, and the statistical average is taken over 4 independent runs. The total number of spins $N$ is related with a linear system size $L$ via $N=3L^2$. By measuring various physical quantities, we identify low-temperature phases.

%parameter dependence at the lowest temperature and explanation of the three spin configurations
Figure \ref{fig:Paradep_structure} (a) shows the parameter dependence of the low-temperature spin structure. There appear three different triple-${\bf Q}$ states as in the isotropic case of $D=0$, the zero-field miniature SkX, collinear, and coplanar states, whose real-space structures are shown in Figs. \ref{fig:Paradep_structure} (b), (c), and (d), respectively. In all the three states, spins residing on each of 3 sublattices at the corners of the small triangle constitute $\uparrow\downarrow\uparrow\downarrow$ chains along the bond directions [see the same colored spins in Figs. \ref{fig:Paradep_structure} (b)-(d)]. The difference among the three consists in the superposition angles of the sublattice $\uparrow\downarrow\uparrow\downarrow$ chains. In the case of Fig. \ref{fig:Paradep_structure} (b), the chains are superposed such that three spins on each triangle be almost orthogonal to one another, forming a miniature skyrmion texture as indicated by a cyan rectangle: outer 4 and inner 1 spins are pointing up and down, respectively, and inbetween, 4 spins form a vortex. Note that all the three states involve 12 spins in their magnetic unit cell.

%the reason for the stability regions
One can see from Fig. \ref{fig:Paradep_structure} (a) that the miniature SkX emerging for $J_1'/J_1 \neq 1$ becomes unstable but is relatively robust against the easy-axis ($D<0$) and easy-plane ($D>0$) magnetic anisotropies. Note that for antiferromagnetic (AFM) $J_1>0$ and ferromagnetic (FM) $J_1'<0$, $|J_1'/J_1|=1$ corresponds to the strongly breathing case with the drastic sign change in the exchange interaction. In the uniform ($J_1'/J_1=1$) and isotropic ($D=0$) case, all the superposition patterns of the $\uparrow\downarrow\uparrow\downarrow$ chains are energetically degenerate, and the collinear configuration is selected by thermal fluctuations \cite{RMO-collinear_Grison_prb_20, KagomeSkX_AK_22}. 
Since the collinear state is compatible with both the easy-axis and easy-plane anisotropies, this state is realized irrespective of the sign of $D$ at $J_1'/J_1=1$. Once the breathing bond-alternation is introduced, the collinear state becomes unstable, and the coplanar state as well as the noncoplanar SkX can be realized at $D=0$ (see Fig. 4 in Ref. \cite{KagomeSkX_AK_22}). In this situation, due to the energetics, the additional easy-axis ($D<0$) and easy-plane ($D>0$) anisotropies favor the collinear and coplanar spin configurations, respectively. As the coplanar state, which is already possible at $D=0$, can be triggered by the magnetic anisotropy more easily than the collinear state, the easy-plane anisotropy is more unfavorable for the miniature SkX than the easy-axis one [see Fig. \ref{fig:Paradep_structure} (a)].

% difference in the ordering properties in the D=0 and D≠0 cases.
Here, we address the difference in the ordering properties between the $D=0$ and $D\neq 0$ cases. In the isotropic Heisenberg case of $D=0$, a long-range magnetic order is not allowed at any finite temperature due to the dimensionality of the system, so that spins still preserve the translational symmetry of the underlying lattice. In this sense, strictly speaking, ''crystal'' of SkX is well-defined only at $T=0$ \cite{KagomeSkX_AK_22}. By contrast, in the anisotropic case of $D\neq 0$, spin components perpendicular (parallel) to the uniaxial direction of the magnetic anisotropy $S_\perp$ ($S_\parallel$) can be quasi-long-range ordered (long-range ordered) with the associated spin correlation length $\xi_{S_\perp}$ ($\xi_{S_\parallel}$) being infinite. Thus, the SkX is well-defined at $T \neq 0$ as well. We note that even for $D=0$, a symmetry breaking of discrete degrees of freedom such as the chirality and a $\mathbb{Z}_2$-vortex topological transition \cite{Z2_Kawamura_84, Sqomega_Okubo_jpsj_10, Z2_AK_prl_20} are possible at finite temperatures \cite{KagomeSkX_AK_22}. 

\begin{figure}
\begin{center}
\includegraphics[scale=0.6]{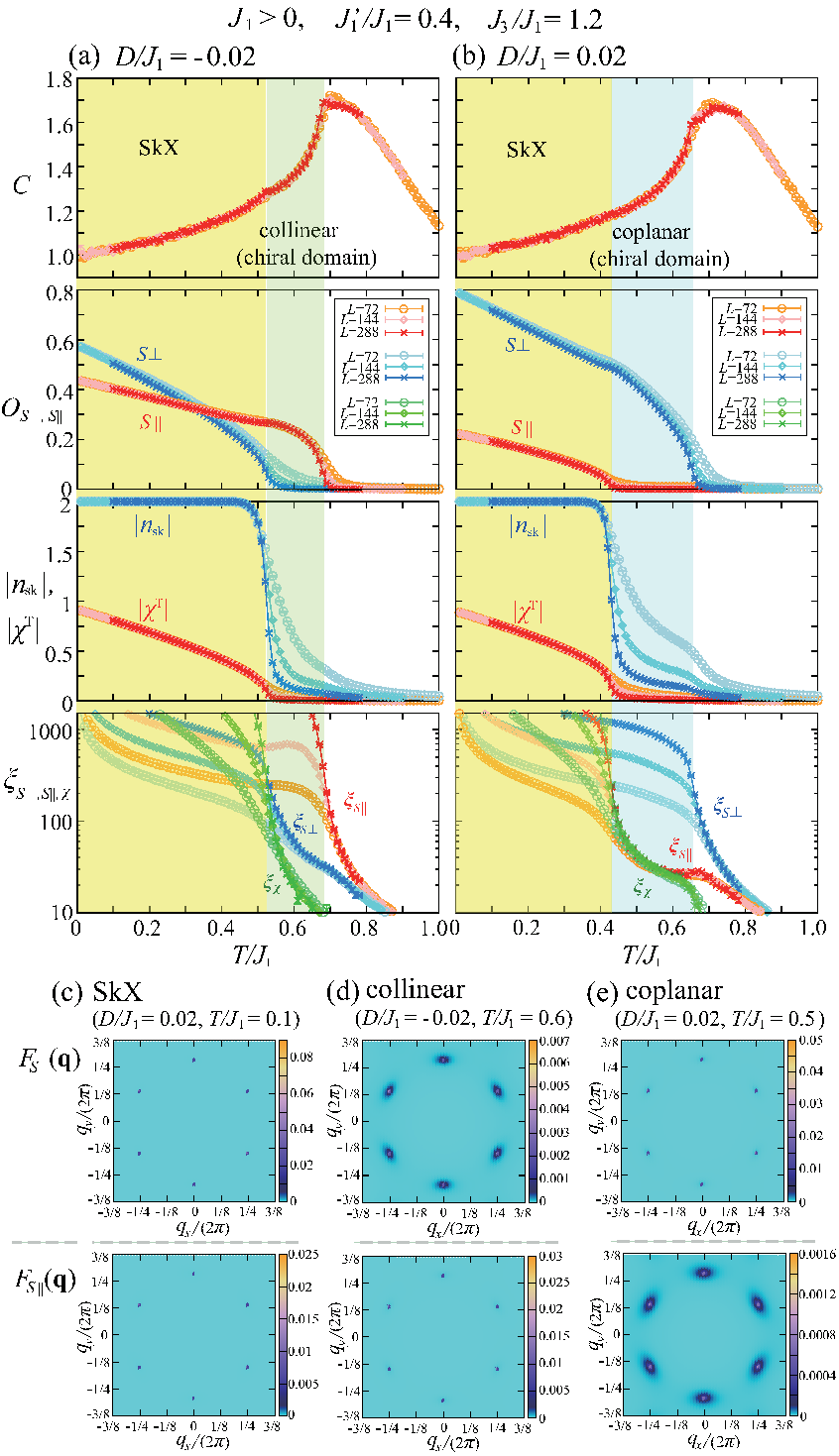}
\caption{(Color online) MC results obtained for $J_1'/J_1=0.4$ and $J_3/J_1=1.2$ with $J_1>0$. (a) [(b)] Temperature dependence of various physical quantities obtained at $D/J_1=-0.02$ ($0.02$), and (c)-(e) the spin structure factors $F_{S_\perp}({\bf q})$ (upper panels) and $F_{S_\parallel}({\bf q})$ (lower panels) in (c) the miniature SkX phase at $D/J_1=0.02$ and $T/J_1=0.1$, (d) the collinear phase at $D/J_1=-0.02$ and $T/J_1=0.6$, and (e) the coplanar phase at $D/J_1=0.02$ and $T/J_1=0.5$, where $S_\perp$ and $S_\parallel$ denote $S^xS^y$ and $S^z$ components of a spin, respectively. Top panels in (a) and (b): the specific heat $C$. The second panels from the top: the averaged intensities of the peaks in $F_{S_\perp}({\bf q})$ and $F_{S_\parallel}({\bf q})$, $O_{S_\perp}$ and $O_{S_\parallel}$. The third panels from the top: the total scalar chirality $|\chi^{\rm T}|$ and the skyrmion number per magnetic unit cell $|n_{\rm sk}|$. Bottom panels: the correlation lengths of the $S_\perp$ and $S_\parallel$ spin components, $\xi_{S_\perp}$ (blue) and $\xi_{S_\parallel}$ (red), and the chiral correlation length $\xi_{\chi}$ (green). 
\label{fig:Tdep} }
\end{center}
\end{figure}

% finite-temperature ordering property
Now, we shall look into the details of phase transitions from the paramagnetic phase, focusing on the AFM $J_1$ and $J_1'$ cases with $J_1'/J_1=0.4$ and $D/J_1=\pm 0.02$, as the associated result for $D=0$ is available for reference \cite{KagomeSkX_AK_22}. To identify the above three phases, we introduce the structure factors for the $S_\perp$ and $S_\parallel$ spin components $F_{S_\perp}({\bf q})=\big\langle|\frac{1}{N}\sum_{\alpha=x,y}\sum_i S_i^\alpha e^{i {\bf q}\cdot{\bf r}_i}|^2\big\rangle$ and $F_{S_\parallel}({\bf q})=\big\langle |\frac{1}{N}\sum_i S_i^z e^{i {\bf q}\cdot{\bf r}_i}|^2 \rangle$.  
As one can see from Figs. \ref{fig:Tdep} (c)-(e), the miniature SkX phase is characterized by quasi-Bragg peaks at the ordering vectors ${\bf Q}_1$, ${\bf Q}_2$, and ${\bf Q}_3$ in $F_{S_\perp}({\bf q})$ [see the upper panel in Fig. \ref{fig:Tdep} (c)] together with Bragg ones in $F_{S_\parallel}({\bf q})$ [see the lower panel in Fig. \ref{fig:Tdep} (c)], whereas as shown in Fig. \ref{fig:Tdep} (d) [(e)], the collinear (coplanar) phase is characterized by the Bragg $F_{S_\parallel}({\bf Q}_\mu)$ [the quasi-Bragg $F_{S_\perp}({\bf Q}_\mu)$] with the counter $S_\perp$ ($S_\parallel$) spin components being short-ranged as indicated by broad peaks in $F_{S_\perp}({\bf q})$ [$F_{S_\parallel}({\bf q})$]. From these peak structures in $F_{S_\perp}({\bf q})$ and/or $F_{S_\parallel}({\bf q})$, all the three phases turn out to be the triple-${\bf Q}$ states.
 
Figures \ref{fig:Tdep} (a) and (b) show the temperature dependence of various physical quantities for the weak easy-axis ($D/J_1=-0.02$) and easy-plane ($D/J_1=0.02$) anisotropies. In the easy-axis case shown in Fig. \ref{fig:Tdep} (a), the system undergoes a phase transition from the paramagnetic phase at $T/J_1 \sim 0.7$, as indicated by a peak in the specific heat $C$ (see the top panel). Below the transition, the $S_\parallel$ Bragg intensity averaged over the three ordering vectors $O_{S_\parallel}=9\sum_{\mu=1,2,3}F_{S_\parallel}({\bf Q}_\mu)/3$ develops (9=3$^2$ is a normalization factor originating from the 3 sublattice), while the $S_\perp$ one $O_{S_\perp}=9\sum_{\mu=1,2,3}F_{S_\perp}({\bf Q}_\mu)/3$ does not (see the second panel from the top), suggesting the occurrence of the collinear phase. At the further low temperature of $T/J_1 \sim 0.5$, $O_{S_\perp}$ and the total scalar chirality $|\chi^{\rm T}|=\langle \frac{1}{2L^2}|\sum_{i,j,k \in \bigtriangleup, \bigtriangledown} \chi_{ijk} | \rangle$ (see the reddish symbols in the third panel from the top) start growing up, pointing to the transition into the SkX phase. Actually, as one can see from the bluish symbols in the third panel from the top in Fig. \ref{fig:Tdep} (a), the skyrmion number per magnetic unit cell $|n_{\rm sk}|=\frac{1}{4\pi}\big\langle  \frac{1}{N/12}|\sum'\Omega_{ijk} | \rangle$ takes the integer value of $|n_{\rm sk}|=2$ down to the lowest temperature, where $\sum '$ denotes the summation over all the triangles that tile up the whole system, and $\Omega_{ijk}$ is evaluated by using spin configurations averaged over 50 MC sweeps to reduce the thermal noise. 

In the case of the weak easy-plane anisotropy shown in Fig. \ref{fig:Tdep} (b), compared with the easy-axis case of Fig. \ref{fig:Tdep} (a), the roles of the $S_\parallel$ and $S_\perp$ spin components are merely interchanged: with decreasing temperature, the $S_\perp$ quasi-Bragg peaks ($O_{S_\perp}$) first develop, and then, the $S_\parallel$ Bragg peaks ($O_{S_\parallel}$) start growing up [see the second panel from the top in Fig. \ref{fig:Tdep} (b)]. The higher and lower temperature phases correspond to the coplanar and SkX states, respectively. Such a difference between the $D<0$ and $D>0$ cases can clearly be seen in the temperature dependence of the spin correlation lengths $\xi_{S_\perp}$ and $\xi_{S_\parallel}$ which are defined by
\begin{equation} 
\xi_{S_{\perp, \parallel}} = \frac{1}{3}\sum_{\mu=1,2,3}\frac{1}{|\mbox{\boldmath $\delta$}_\mu|} \sqrt{\frac{F_{S_{\perp,\parallel}}({\bf Q}_\mu)}{F_{S_{\perp,\parallel}}({\bf Q}_\mu+\mbox{\boldmath $\delta$}_\mu)}-1 },
\end{equation}
with $\mbox{\boldmath $\delta$}_\mu=\frac{2\pi}{\sqrt{3} \, L}\hat{Q}_\mu$. We note that the peak width of $F_{S_{\perp,\parallel}}({\bf Q}_\mu)$ is anisotropic in the $q_x$-$q_y$ plane [see, for example, the upper panel in Fig. \ref{fig:Tdep} (d)]. Although we have taken $\mbox{\boldmath $\delta$}_\mu$ along the shortest direction in the elliptical-shaped peak tail, the choice of $\mbox{\boldmath $\delta$}_\mu$ does not affect the following result qualitatively. As readily seen from the bottom panels in Figs. \ref{fig:Tdep} (a) and (b), $\xi_{S_\perp}$ and $\xi_{S_\parallel}$ tend to diverge at different transition temperatures. In addition to the spin sector, we also calculate a correlation length in the chiral sector, i.e., the chiral correlation length
\begin{equation} 
\xi_{\chi} = \frac{1}{3}\sum_{\mu=1,2,3}\frac{1}{|\mbox{\boldmath $\delta$}_\mu|} \sqrt{\frac{F_{\chi}(0)}{F_{\chi}(\mbox{\boldmath $\delta$}_\mu)}-1 }
\end{equation}
with $F_{\chi}({\bf q})=\langle \frac{1}{L^2}|\sum_{i,j,k \in \bigtriangleup} \chi_{ijk} e^{i{\bf q}\cdot{\bf r}_\bigtriangleup} |^2 \rangle$ and ${\bf r}_\bigtriangleup$ being the center of mass position of a small triangle $i, \, j,$ and $k$.
One can see from the bottom panel in Fig. \ref{fig:Tdep} (a) [(b)] that $\xi_\chi$ gradually increases to diverge at the transition into the SkX phase with $|\chi^T|\neq 0$, similarly to $\xi_{S_\perp}$ ($\xi_{S_\parallel}$). Interestingly, in the $D<0$ case of Fig. \ref{fig:Tdep} (b), $\xi_\chi$ almost coincides with $\xi_{S_\parallel}$ in the coplanar phase, which might be due to the fact that both the chirality $\chi_{ijk}$ and $S_\parallel$ spin component are discrete $\mathbb{Z}_2$ degrees of freedom. We note that for strong anisotropies, the SkX state is not realized [see Fig. \ref{fig:Paradep_structure} (a)], and thereby, $\xi_\chi$ remains very short. As will be explained below, the development of $\xi_\chi$ toward the SkX phase is reflected in the real-space structure as a chiral-domain growth.  

\begin{figure}
\begin{center}
\includegraphics[scale=0.83]{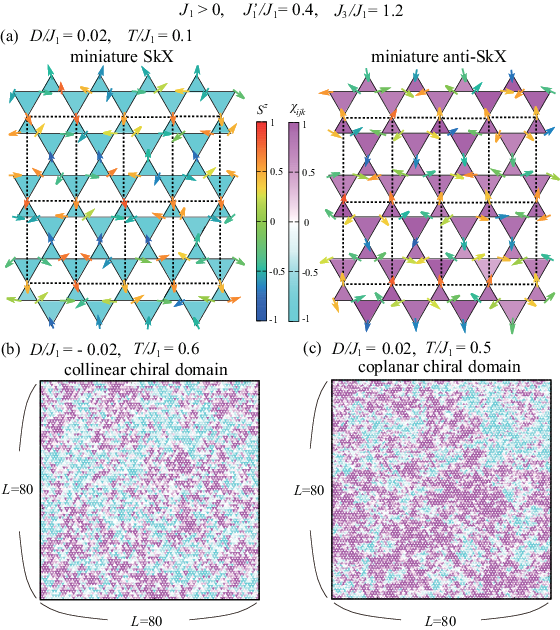}
\caption{(Color online) MC snapshots of the chirality distributions in the (a) low-temperature SkX/anti-SkX, (b) higher-temperature collinear, and (c) higher-temperature coplanar phases, where the system parameters are the same as those in Figs. \ref{fig:Tdep} (c)-(e). In (a), the left (right) panel shows the SkX (anti-SkX) structure with negative (positive) total chirality $\chi^T$, where on each triangle, the color represents the local scalar chirality $\chi_{ijk}$, and an arrow and its color represent the $S^xS^y$ and $S^z$ components of a spin, respectively. A unit cell of the skyrmion is indicated by a dotted rectangle. In (b) and (c), positive- and negative-chirality domains are randomly distributed.  
\label{fig:chi_dis} }
\end{center}
\end{figure}

Figure \ref{fig:chi_dis} (b) [(c)] shows the real-space distribution in the collinear phase at $T/J_1=0.6$ (coplanar phase at $T/J_1=0.5$) just above the transition into the SkX phase. Reflecting the fact that in the lower-temperature SkX phase, the SkX and anti-SkX each having negative or positive uniform chirality are energetically degenerate [see Fig. \ref{fig:chi_dis} (a)], the higher-temperature collinear and coplanar phases possess random domain structures consisting of positive- and negative-chirality clusters. Thus, these collinear and coplanar phases appearing between the SkX and paramagnetic phases can be viewed as a chiral domain state. In the cases of Figs. \ref{fig:chi_dis} (b) and (c), the associated $\xi_\chi$'s at $T/J_1=0.6$ and $0.5$ are roughly 20 and 40 lattice spacings, respectively [see the bottom panels in Figs. \ref{fig:Tdep} (a) and (b)], which turn out to be of the same order as the linear domain size. Compared with the field-induced chiral domain state, i.e., the $Z$ phase \cite{SkX_Okubo_12, SkX-RKKY_Mitsumoto_21, SkX-RKKY_Mitsumoto_22}, the present spin-collinear chiral-domain state has the same ordering properties as those of the $Z$ phase: $\xi_{S_\parallel}$ is infinite, while $\xi_{S_\perp} \sim \xi_\chi$ is relatively long but finite. In the spin-coplanar chiral-domain state, the roles of $\xi_{S_\parallel}$ and $\xi_{S_\perp}$ are interchanged, but there is no significant difference in the chiral sector. 

Finally, we will address experimental implications of our result. In the chiral domain states, the positive and negative chiralities are canceled out, so that the topological Hall effect originating from the total chirality is absent, which is in sharp contrast to the miniature SkX phase with $|\chi^{\rm T}| \neq 0$ where the topological Hall effect is possible even at zero field, taking, in principle, either of the positive and negative signs depending on experimental conditions. The fundamental ingredients for the zero-field miniature SkX are the strong AFM $J_3$ along the bond direction and the breathing bond-alternation. Although so-far reported breathing-kagome magnets do not possess a strong $J_3$ \cite{GdRuAl_Hirschberger_natcom_19, DQVOF_Orain_prl_17, DQVOF_Clark_prl_13, LiAMoO_Haraguchi_prb_15, LiAMoO_Sharbaf_prl_18, PbOFCu_Zhang_ChemComm_20, YbNiGe_Takahashi_jpsj_20, FeSn_Tanaka_prb_20, CaCrO_Balz_natphys_16}, the uniform kagome antiferromagnet BaCu$_3$V$_2$O$_8$(OD)$_2$ \cite{CoplanarOct_Boldrin_prl_18} possesses a considerablly strong $J_3$ and its low-temperature ordered phase accompanied by tiny lattice distortions is the 12-sublattice coplanar state, the same spin structure as that shown in Fig. \ref{fig:Paradep_structure} (c), pointing to the possibility of the zero-field miniature SkX in its family compounds. Also, the understanding of the origin of such a strong AFM $J_3$ would help the exploration of various types of topological spin textures such as the hedgehog lattice emerging in $J_1$-$J_3$ breathing-pyrochlore antiferromagnets \cite{hedgehog_AK_prb_21, hedgehog_AK_prb_22}. 

\begin{acknowledgments}
We are thankful to ISSP, the University of Tokyo and YITP, Kyoto University for providing us with CPU time. This work is supported by JSPS KAKENHI Grant Number JP21K03469.
\end{acknowledgments}

\end{document}